\documentclass[12pt]{article}

\font\sln=cmfi10 at 11pt
\font\sll=cmfi10 at 12pt
\font\sls=cmfi10 at 9pt

\def\i{\mbox{i}}
\def\d{\mbox{d}}
\def\Im{\mbox{Im}}
\def\Re{\mbox{Re}}
\def\G{\mbox{\sln\char'000}}
\def\GG{\mbox{\sll\char'000}}
\def\g{\mbox{\sls\char'000}}

\title{
$\overline{\mbox{OMS}}$ scheme of UV renormalization \\
in the presence of unstable fundamental particles}
\author{M. L. Nekrasov \\
{\small Institute for High Energy Physics, 142284 Protvino,
Russia}}

\date{}

\begin{document}

\maketitle

\begin{abstract}
\noindent A generalization of the on-mass-shell scheme of UV
renormalization (the $\overline{\mbox{OMS}}$ scheme) to the case
of presence of unstable fundamental particles is proposed. Its
basic ingredients are as follows: (i) the renormalized mass
coincides with a real part of the position of the complex pole of
the corresponding propagator, (ii) the imaginary part of the
on-shell self-energy coincides with the imaginary part of the
complex pole position. The latter property implies the
gauge-invariance of the imaginary part of the on-shell self-energy
in the $\overline{\mbox{OMS}}$ scheme and its direct connection
with the width of the unstable particle. Starting with the
three-loops this connection becomes nontrivial.
\end{abstract}

\bigskip

The aim of this paper is to introduce an effective generalization
of the on-mass-shell (OMS) scheme of UV renormalization to the
case of presence of unstable fundamental particles. This problem
is determined by the difficulties with the gauge invariance, noted
in the framework of the conventional generalization of the OMS
scheme \cite{Veltman,Denner,B-P} in the cases of W, Z and Higgs
bosons beyond the one-loop order
\cite{Sirlin1,Sirlin2,SirlinW,SirlinH}. In fact, however, even in
the case of non-gauge field theories the conventional
generalization \cite{Veltman,Denner,B-P} ceases to have those
attractive properties which are peculiar to the standard OMS
scheme in the case of stable particles. So, finding the ``true''
generalization of the OMS scheme, possessing the
physically-motivated (and, hence, convenient) properties, is an
important task from the general field-theoretic point of view.

Let us begin our analysis with considering the inverse
renormalized propagator of a scalar particle, or of
$\delta_{\mu\nu}$-part of a vector particle. We do not define
precisely the sort of particle and the underlying theory since the
problem of renormalization is general enough in nature. In terms
of the renormalized quantities we have
\begin{equation}\label{1}
\Delta^{-1}(s) = s - M^2 - \delta Z (s-M^2) - \delta M^2 +
\Sigma(s)\,.
\end{equation}
Here $M^2$ is the renormalized lagrangian mass, $\Sigma(s)$ is the
self-energy that depends, besides $s$, also on $M^2$ and the
renormalized coupling constant $\alpha$. Quantities $\delta M^2$
and $\delta Z$ describe the counterterm contributions (notice the
minus sign in $\delta Z$ in our notation). Their assigning is to
cancel UV divergencies in $\Sigma(s)$.

In the framework of perturbation theory this cancellation should
be performed order-by-order. So, with
\begin{equation}\label{2}
\Sigma(s) = \sum_{n=1}^{\infty} \alpha^n \Sigma_n(s) \,,
\end{equation}
\vspace*{-1\baselineskip}
\begin{equation}\label{3}
\delta Z = \sum_{n=1}^{\infty} \alpha^n \, C^{Z}_n\,,  \quad
\delta M^2 = \sum_{n=1}^{\infty} \alpha^n \, C^{M}_n\,,
\end{equation}
the coefficients $C^{Z}_n$ and $C^{M}_n$ must provide finiteness
for $\Sigma_n(s) - C^M_n\! - C^Z_n (s-M^2)$. From the unitarity of
the $S$-matrix it follows that the counterterms must be real
\cite{Bogol-Shirk}. The operational use of various renormalization
schemes confirms that in the commonly used (gauge) theories two
real counterterms indeed cancel UV divergences in $\Sigma_n(s)$.

It is worth reminding that various renormalization schemes are
different in finite parts of counterterms. This difference, in
turn, means a different determination of the renormalized
lagrangian parameters and the normalization of the Green
functions. In the standard OMS scheme the renormalized mass $M^2$
is made equal to the physical mass $M^2_{\mbox{\scriptsize Ph}}$
determined by $\Delta^{-1}(M^2_{\mbox{\scriptsize Ph}}) = 0$.
Besides, the residue at the pole in the propagator is made equal
to~1. Both these properties make the OMS scheme very convenient
for the practical usage.

In the case of stable particles the above-mentioned properties are
provided by the following counterterms:
\begin{equation}\label{4}
C^M_n = \Sigma_n(M^2), \quad C^Z_n = \Sigma'_n(M^2)\,.
\end{equation}
However, when the particle under consideration is unstable, this
choice of counterterms is not admissible because of the
non-vanishing imaginary parts in the self-energy.

The most commonly used way \cite{Veltman,Denner,B-P} of solving
the problem consists in replacing (4) by
\begin{equation}\label{5}
C^M_n = \Re\,\Sigma_n(M^2), \quad C^Z_n = \Re\,\Sigma'_n(M^2)\,.
\end{equation}
However, then the renormalized mass becomes defined by the
condition $\Re\,\Delta^{-1}(M^2)=0$, which does not provide the
pole to the propagator. As a result, the renormalized mass becomes
no longer physical observable. In the case of electroweak theory
this fact manifests itself in the emergence of the
gauge-dependence in the renormalized masses of the vector bosons
and the Higgs boson \cite{Sirlin1,Sirlin2,SirlinW,SirlinH}. This
situation is objectionable and certainly must be cured in a true
generalization of the OMS scheme.

Actually, the latter problem has been posed not once. The idea of
its solution consists in equating the renormalized mass $M^2$ to a
real part of the position of the complex pole $s_{p}$ of the
propagator, which is gauge-invariant
\cite{Sirlin1,Sirlin2,SirlinW,SirlinH,Stuart}. In
Ref.~\cite{2-loop} this idea has been implemented in a special
case of calculation of the two-loop correction to the muon
lifetime. However, the general study of the problem has not been
made. So, the true generalization of the OMS scheme is still not
completed. In particular, the second renormalization condition for
$\Sigma'(s)$ is still not determined properly. The point is that
the non-vanishing $\Im\,\Sigma'(s)$ prevents the residue in the
pole from being equal to~1. In Ref.~\cite{2-loop} the second
renormalization condition was chosen rather formally, in the form
of (5). In the particular case of the two-loop calculation of the
muon lifetime this choice did not have adverse consequences.
However, on description of the production and decay of unstable
particles this choice may lead again to difficulties with gauge
invariance (see below).

In the present paper we propose an unconventional way of fixing
the second renormalization condition. It has a clear physical
significance, so the name ``physical'' can be appropriated to this
scheme. We call it the $\overline{\mbox{OMS}}$ scheme. Under the
limit of switching-off the instability, it transforms smoothly to
the standard OMS scheme.

The basic point of our consideration is the condition of the
gauge-invariance of the position of the complex pole $s_{p}$
\cite{Sirlin1,Sirlin2,SirlinW,SirlinH,Stuart,Gambino}. Owing to
(1) the equation for $s_{p}$, which is $\Delta^{-1}(s_{p}) = 0$,
may be rewritten in the form
\begin{equation}\label{6}
s_{p} = M^2 + \delta M^2 + \delta Z (s_{p}-M^2) - \Sigma(s_{p})\,.
\end{equation}
With the aid of (2) and (3) this equation can be solved by
iteration. So, denoting the solution up to $O(\alpha^{n+1})$
correction by $s_{pn}$, and introducing the short-card notation
$R_n=\Re\,\Sigma_n(M^2)$, $I_n=\Im\,\Sigma_n(M^2)$, with the
primed symbols indicating the derivatives, we get the following
sequence of iterative solutions:
\begin{eqnarray}\label{7}
s_{p0} &=& M^2 \,,
                                        \\[0.2\baselineskip]
s_{p1} &=& s_{p0} \:
        + \alpha   (C^M_1\! - R_1 - \i I_1)\,,
                                        \\[0.2\baselineskip]
s_{p2} &=& s_{p1} \:
        + \alpha   (s_{p1}-s_{p0})(C^Z_1\!-R'_1-\i I'_1)
                               \nonumber\\[0.2\baselineskip]
       & & \quad\;\:\,
        + \, \alpha^2 (C^M_2\!-R_2-\i I_2)\,,
                                        \\[0.2\baselineskip]
s_{p3} &=& s_{p2} \:
        + \alpha   (s_{p2}\!-\!s_{p1})(C^Z_1\!-R'_1-\i I'_1)
                               \nonumber\\[0.2\baselineskip]
       & & \quad\;\:\,
        + \,{\scriptstyle\frac{1}{2}}
          \alpha (s_{p1}\!-\!s_{p0})^2 (-\,R''_1-\i I''_1)
                               \nonumber\\[0.2\baselineskip]
       & & \quad\;\:\,
        + \,\alpha^2 (s_{p1}\!-\!s_{p0})(C^Z_2\!-R'_2-\i I'_2)
                               \nonumber\\[0.2\baselineskip]
       & & \quad\;\:\,
        + \,\alpha^3 (C^M_3\!-R_3-\i I_3)\,,
                                        \\
\cdots & & \cdots \nonumber
\end{eqnarray}

For methodological reasons we consider, at first, the
conventionally generalized OMS scheme \cite{Veltman,Denner,B-P}
determined by (5). Then, the listed above solutions are reduced to
\begin{eqnarray}\label{11}
s_{p0} &=& M^2 \,,\\[0.2\baselineskip] s_{p1} &=& M^2\! - \i
\alpha I_1\,,
\\[0.2\baselineskip]
s_{p2} &=& M^2\! - \alpha^2 I_1 I'_1 - \i \alpha I_1
               -\i \alpha^2 I_2\,,
\\[0.2\baselineskip]
s_{p3} &=& M^2\! - \alpha^2 I_1 I'_1
               - \alpha^3 (I_1 I'_2 + I'_1 I_2 -
               {\scriptstyle\frac{1}{2}}I^2_1 R''_1)
                               \nonumber\\[0.2\baselineskip]
       & &     - \,\i \alpha I_1 -\i \alpha^2 I_2 - \i \alpha^3
[I_3 - I_1 (I'_1)^2 - {\scriptstyle\frac{1}{2}}I^2_1 I''_1],
                                        \\
\cdots & & \cdots \nonumber
\end{eqnarray}
From formulas (11) and (12) we see that in the case of gauge
theories the renormalized mass $M^2$ is gauge-invariant up to
$O(\alpha^2)$ correction. However, the $O(\alpha^2)$ correction is
gauge-dependent since the difference $M^2\!-\Re\,s_{p2} = \alpha^2
I_1 I'_1$ is like that. This property follows from the
gauge-invariance of $I_1$, which is the consequence of (12), and
the gauge-dependence of $I'_1$. The latter property was observed
in the case of Z-boson \cite{Sirlin1,Sirlin2}, W-boson
\cite{SirlinW}, and Higgs boson \cite{SirlinH}. So, the
gauge-invariance of $s_{p}$ implies the gauge-dependence of the
renormalized mass $M^2$ at the two-loop order \cite{PT}.

It should be noted that from the viewpoint of underlying
principles there is nothing catastrophic in the latter situation,
since the renormalized mass is not an observable quantity.
However, it is not reasonable to use in practice such
renormalization scheme. A better choice is a scheme where the
renormalized mass is gauge-invariant, and it would be even better
if the renormalized mass coincided with the observable
$\Re\,s_{p}$.

Now we proceed directly to the construction of the
$\overline{\mbox{OMS}}$ scheme, paying special attention to the
choice of the second renormalization condition. We do that in an
iterative manner, order-by-order. So, in the leading order we have
$s_{p0}=M^2$ without alternatives. In the one-loop order we set
\begin{equation}\label{15}
C^M_1 = R_1 \,.
\end{equation}
Then, $s_{p1}$ coincides with that of formula (12).

The difference with the conventionally generalized OMS scheme
\cite{Veltman,Denner,B-P} appears starting with the two-loop
order. Owing to (8), (9) and (15), we have
\begin{eqnarray}\label{16}
&s_{p2} =
 M^2 - \i \alpha I_1 - \i \alpha^2 I_1 (C^Z_1\!-R'_1-\i I'_1)&
                               \nonumber\\[0.2\baselineskip]
         & \;\;\: + \alpha^2 (C^M_2\!-R_2-\i I_2) \,.&
\end{eqnarray}
By assuming,
\begin{equation}\label{17}
C^Z_1 = R'_1 \,,
\end{equation}
we come to the same imaginary part in $s_{p2}$ as in (13).
However, in order to satisfy requirement $\Re\,s_{p} = M^2$, we
have to impose a different condition for $C^M_2$ (cf.
\cite{2-loop}):
\begin{equation}\label{18}
C^M_2 = R_2 + I_1 I'_1 \,.
\end{equation}
So, taking into account (17) and (18), we obtain
\begin{equation}\label{19}
s_{p2} = M^2 - \i \alpha I_1 - \i \alpha^2 I_2 \,.
\end{equation}

The difference becomes more considerable in the three-loop order.
Owing to (10), (12), (17) and (19), we have
\begin{eqnarray}\label{20}
s_{p3} &=& s_{p2} - \i \alpha^3 I_3
        - \i \alpha^3 I_1 (C^Z_2\!-R'_2
        - {\scriptstyle\frac{1}{2}}\,I_1 I''_1)
                               \nonumber\\[0.2\baselineskip]
       & & \quad\;\:
        + \,\alpha^3 (C^M_3\!-R_3 - I_2 I'_1 - I_1 I'_2
        + {\scriptstyle\frac{1}{2}}\,I_1^2 R''_1) \,.
\end{eqnarray}
Let us note, that the imaginary part of $s_{p3}$ has a far
complicated structure. However, by assuming
\begin{equation}\label{21}
C^Z_2 = R'_2 + {\scriptstyle\frac{1}{2}}\,I_1 I''_1\,,
\end{equation}
we can get the simplest possible expression for $\Im\,s_{p3}$,
namely $-\i\alpha^3 I_3$. In order to provide $\Re\,s_{p} = M^2$,
we set
\begin{equation}\label{22}
C^M_3 = R_3 + I_2 I'_1 + I_1 I'_2 -
{\scriptstyle\frac{1}{2}}\,I_1^2 R''_1 \,.
\end{equation}
As a result, we come to
\begin{equation}\label{23}
s_{p3} = M^2 - \i \alpha I_1 - \i \alpha^2 I_2 - \i \alpha^3 I_3
\,.
\end{equation}

The above consideration may be continued up to any~$n$, providing
in the limit $n \to \infty$ the following solution \cite{Sirlin3}:
\begin{equation}\label{24}
s_{p} = M^2 - \i \, \Im \Sigma(M^2) \,.
\end{equation}

Let us summarize the main features of the above construction. At
any step $n$, when considering the imaginary part of $s_{pn}$, we
fix the renormalization condition for $C^Z_{n-1}$ by imposing the
requirement $\Im(s_{pn}\!-\!s_{p(n\!-\!1)})\!=\!-\alpha^n I_n$.
When considering the real part of $s_{pn}$, we fix the
renormalization condition for $C^M_{n}$ by imposing
$\Re\,s_{pn}\!=\!M^2$. The divergent contributions to $C^M_{n}$
and $C^Z_{n-1}$ turn out to be $R_{n}$ and $R'_{n-1}$,
respectively. The resulting formulas for $C^M_{n}$ and $C^Z_{n-1}$
can be obtained for any $n$. However, the cases with $n\!\ge\!4$,
most likely, will not be claimed in a foreseeable future. So, we
will not be wasting time to find the general solution.

Let us turn now to the discussion.

{\bf 1}. The first question is about the structure of the
propagator in the resonance region. By excluding $\delta M^2$ from
(6) in favor of $s_{p}$, one can derive from (1),
\begin{eqnarray}\label{25}
\Delta^{-1}(s) &=& (s\!-\!s_{p}) (1 - \delta Z)
                + \Sigma(s) - \Sigma(s_{p})
                \nonumber\\[0.2\baselineskip]
               &=& (s\!-\!s_{p})
               \left[1 - \delta Z + \Sigma'(s_{p})\right]
                + O\left((s\!-\!s_{p})^2\right).
\end{eqnarray}
From (25) we see that the renormalized propagator has a complex
pole with the residue free from UV divergences \cite{IR}. The
latter property follows from the fact that the difference
$\Sigma'(s_{p}) - \delta Z$ is finite, because the UV divergence
in $\Sigma'(s_{p})$ is equivalent to that in $\Re\Sigma'(M^2)$ and
the latter one is cancelled by $\delta Z$ in any scheme. However,
in the unstable-particle case, in view of non-zero
$\Im\Sigma'(s_{p})$, there is no way to make the residue equal
to~1. Moreover, in most cases the real part in the residue is not
equal to~1, either. For instance, in the generalized by
\cite{Veltman,Denner,B-P,2-loop} OMS schemes, where the second
renormalization condition is determined by the second formula in
(5), one has $1 - \delta Z + \Sigma'(s_{p}) = 1 + \i \alpha I'_1 +
\alpha^2 I_1 I''_1 + \i \alpha^2 (I'_2 - I_1 R''_1) +
O(\alpha^3)$. In the $\overline{\mbox{OMS}}$ scheme, $1 - \delta Z
+ \Sigma'(s_{p}) = 1 + \i \alpha I'_1 + \frac{1}{2}\alpha^2 I_1
I''_1 + \i \alpha^2 (I'_2 - I_1 R''_1) + O(\alpha^3)$.

{\bf 2}. The second point concerns the renormalization of the
coupling constants. Formally, the renormalization prescription for
coupling constants is imposed separately from that for
propagators. In the electroweak theory it may be the same as in
the conventionally generalized OMS scheme \cite{B-P}. Namely, the
$U(1)$ constant $e^2$ can be determined as the electric charge
measured by the Compton process at the low-energy limit. The weak
mixing and the weak coupling constant can be determined by
relations $s^2_{W}=1-M^2_W/M^2_Z$ and $g^2=e^2/s^2_{W}$, which are
considered to be valid in all the orders of perturbation theory.
It should be noted, however, that the actual renormalization of
the coupling is determined not only by the renormalization of the
coupling constant, but also by the wave-function renormalization
constants of the particles participating in the interaction. So,
the actual renormalized couplings, starting with the two-loop
order, become different in the generalized by
\cite{Veltman,Denner,B-P,2-loop} OMS schemes and in the
$\overline{\mbox{OMS}}$ scheme.

{\bf 3}. In gauge theories considered in the framework of the
renormalization scheme with the gauge-invariant renormalized
masses, there is an additional constraint on the counterterms
following from the gauge-invariance of bare masses. Really, the
bare mass connects with the renormalized mass by means of the
relation
\begin{equation}\label{26}
M^2_0\!=\!M^2 + (1-\delta Z)^{-1}\delta M^2\,.
\end{equation}
So, from the gauge-invariance of $M^2_0$ and $M^2$ the
gauge-invariance of $(1-\delta Z)^{-1}\delta M^2$ follows. At the
one-loop order this condition implies the gauge-invariance of $R_1
\equiv \Re\Sigma_1(M^2)$. Notice, due to the gauge-invariance of
$M^2$ at the one-loop order, this particular corollary is common
for all the above-considered versions of the generalized OMS
schemes. In case of unstable bosons in the electroweak theory this
property was independently noted on the base of direct
calculations \cite{B-P} (it was the consequence of the consistent
taking into account the tadpole contributions). At the two-loop
order, in the generalized by \cite{2-loop} OMS scheme and in the
$\overline{\mbox{OMS}}$ scheme, the above condition implies the
gauge-invariance of $R_2 + R_1 R'_1 + I_1 I'_1$. At the higher
orders the corresponding constraints in these schemes become
different.

{\bf 4}. In some cases the $\overline{\mbox{OMS}}$ scheme is
preferable with respect to the OMS scheme generalized in the sense
of \cite{2-loop}. For instance, this is the case with
unstable-particle production and decay within the two-loop
precision. Really, in view of (25), the propagator in the
resonance region, $s-M^2=O(\alpha)$, within this precision may be
approximated by the expression
\begin{eqnarray}\label{27}
&\Delta^{-1}(s) \simeq (s-s_{p3}) \left[ 1 + \frac{1}{2}\alpha
(s-s_{p3}) R''_1 \right] (\mbox{Res})^{-1},&
\end{eqnarray}
where $\mbox{Res}\!=\![1-\delta Z+\Sigma'(s_{p})]^{-1}$ is the
residue in the pole (see the foregoing formulas in different
schemes), and $s_{p3}$ is the pole within the three-loop
precision. In the $\overline{\mbox{OMS}}$ scheme $s_{p3}$ is
determined by (23), while in the generalized by \cite{2-loop} OMS
scheme it is determined~by
\begin{eqnarray}\label{28}
&s_{p3} = M^2 - \i \alpha I_1 - \i \alpha^2 I_2 - \i \alpha^3 (I_3
- \frac{1}{2}\,I_1^2 I''_1)\,.&
\end{eqnarray}
Note, in both cases $s_{p3}$ includes the $I_3$ contribution. The
common practice of taking into account the imaginary contribution
to self-energy is via the unitarity relation, which relates it to
the width of unstable particle at the less-loop order. However,
while the width is always gauge-invariant, the imaginary part in
self-energy is not always that. Really, in the generalized by
\cite{2-loop} OMS scheme $I_3$ is gauge-dependent, which is seen
from (28) and the gauge-dependence of $I''_1$. At the same time,
in the $\overline{\mbox{OMS}}$ scheme $I_3$ is gauge-invariant.
So, in the $\overline{\mbox{OMS}}$ scheme $I_3$ can directly be
related to the width of unstable particle, but not in the
generalized by \cite{2-loop} OMS scheme.

{\bf 5}. The above-mentioned relation may be derived from the
formula for the lifetime of an unstable particle. Below, pursuing
the illustrative purposes, we present rather a heuristic
derivation of this formula. So, in as much as possible idealized
statement of the problem, the lifetime is directly connected with
the propagator of unstable particle. Really, the amplitude of
production of unstable particle (anywhere in the Universe) and its
subsequent decay after the time $x^0$, is proportional to
\begin{equation}\label{29}
A(x^0) \sim \! \int \! \d \vec x \!
\int \!\! \frac{\d p}{(2\pi)^4} \: e^{- \mbox{\scriptsize i} p x}
\Delta(p^2)
  = \! \int \! \frac{\d E}{2\pi}\:
    e^{- \mbox{\scriptsize i} E x^0} \! \Delta(E^2)\,.
\end{equation}
The remaining integral can be calculated with the aid of (25). By
assuming parametrization $\Im\,s_{p} = M \GG$, we get
\begin{equation}\label{30}
A(x^0) \sim e^{- \mbox{\scriptsize i}
               x^0 M \sqrt{1 - \mbox{\scriptsize i} \,\g /M}}\,.
\end{equation}
Then, the normalized-to-one probability is
\begin{equation}\label{31}
P(x^0) = \frac{|A(x^0)|^2}{\int_0^{\infty} \d x^0 |A(x^0)|^2} =
\frac{1}{T} \, e^{-x^0/T} \,,
\end{equation}
with $T$ is the lifetime. The direct calculation gives
\begin{eqnarray}\label{32}
\frac{1}{T} = &M \sqrt{2\!\left( \sqrt{1+\frac{\G^{2}}{M^2} } - 1
\right)}& = \G - \frac{\G^{3}}{8M^2} + \cdots \,,
\end{eqnarray}
with dots standing for $O(\GG^{5}/M^4)$ correction. By identifying
$T^{-1}$ with the width $\Gamma$ of unstable particle, we derive
from (32) and (24) the formula
\begin{equation}\label{33}
I_3 =  M \Gamma_{\mbox{\scriptsize 2-loop}} +
\Gamma^3_{\mbox{\scriptsize 0-loop}} \left/ (8\,M^2)\right.,
\end{equation}
which is valid in the $\overline{\mbox{OMS}}$ scheme only. The
origin of the second term in (33) may be associated with the
triple cut emerging while applying the Cutkosky rules at the
three-loop level.

In summary, we have constructed the $\overline{\mbox{OMS}}$ scheme
of UV renormalization, which we consider as most suitable for
applications with unstable particles. Really, the renormalized
mass in this scheme coincides with the physical mass of unstable
particle, and the on-shell self-energy is directly connected with
its width. Both these quantities are the observables. So, the
$\overline{\mbox{OMS}}$ scheme absorbs all the conveniences of the
well-known complex pole scheme \cite{Stuart} for the
parametrization of the amplitude.

The practical significance of the $\overline{\mbox{OMS}}$ scheme
is obvious in the case of the processes of unstable-particle
production and decay considered with the two-loop (and higher)
precision. Such processes are to be studied at the future
colliders \cite{Future}.

{\it Acknowledgements.} The author is grateful to D.Bardin for
useful discussions, for indication to Ref.\cite{2-loop}, and for a
kindly presented opportunity of a detailed acquaintance with his
and G.Passarino book \cite{B-P} and lectures \cite{Bardin}. The
author thanks A.Sirlin for pointing out on Ref.\cite{Sirlin3} and
on misprints in Eqs.(32) and (33) in the initial version of the
paper.


\begin{thebibliography}{99}
\bibitem{IR}
We do not consider the IR divergences arising in the presence of
massless particles. Such divergences are to be eliminated by
special methods.
\end{thebibliography}
\end{document}